\documentclass[aps,pre,twocolumn,showpacs,showkey,square,numbers,amssymb,amsmath,nobibnotes,mathtools,longbibliography]{revtex4-1}
\usepackage{bm}
\usepackage{times,float}
\usepackage{graphicx}
\usepackage[usenames,dvipsnames,svgnames]{xcolor}
\usepackage{hyperref}
\hypersetup{colorlinks=true, linkcolor=NavyBlue, citecolor=PineGreen,urlcolor=cyan}

\usepackage{color}
\definecolor{dgreen}{rgb}{0,0.7,0}

\begin{document}
\title{Nonequilibrium dynamics of the Ising model on heterogeneous networks with an arbitrary distribution of threshold noise}
\author{Leonardo S. Ferreira}
\author{Fernando L. Metz}
\address{Physics Institute, Federal University of Rio Grande do Sul, 91501-970 Porto Alegre, Brazil}

\begin{abstract}
  The Ising model on networks plays a fundamental role as a testing ground for understanding cooperative phenomena in complex systems.
  Here we solve the synchronous dynamics of the Ising model on random graphs with an arbitrary degree distribution in the high-connectivity limit.
  Depending on the distribution of the threshold noise that governs the microscopic dynamics, the model evolves to nonequilibrium
  stationary states.
  We obtain an exact dynamical equation
  for the distribution of local magnetizations, from which we find the critical line that separates the paramagnetic from the ferromagnetic phase.
  For random graphs with a negative binomial degree distribution, we demonstrate that the stationary critical behavior as well as the long-time critical dynamics of the first two moments
  of the local magnetizations depend on the distribution of the threshold noise.
  In particular, for an algebraic threshold noise, these critical properties are determined by the power-law tails of the distribution of thresholds.
  We further show that
  the relaxation time of the average magnetization inside each phase exhibits the standard mean-field critical scaling.
  The values of all critical exponents considered here are independent of the variance of the negative binomial degree distribution.
  Our work highlights the importance
  of certain details of the microscopic dynamics for the critical behavior of nonequilibrium spin systems.
\end{abstract}
%\pacs{02.50.−r,05.70.Fh,02.10.Ox}
\maketitle
  
%%%%%%%%%%%%%%%%%%%%%%%%%%%%%%%%%%%%%%%%%%%%%%%%%%%%%%%%%%%%%%%%%%%%%%%%%%%%%%%%%%%%%%%%%%%%%%
%%%%%%%%%%%%%%%%%%%%%%%%%%%%%%%%%%%%%%%%%%%%%%%%%%%%%%%%%%%%%%%%%%%%%%%%%%%%%%%%%%%%%%%%%%%%

\section{Introduction}
\label{sec:intro}

Understanding cooperative phenomena in large interacting complex systems is at the forefront of various 
branches of science \cite{Doro2008,Barrat2008,Castellano2009,Mezard2009}.
The Ising model on random graphs provides a general framework to tackle this problem and to 
explore how heterogeneous interactions among the spins influence their dynamical behavior.
Heterogeneity here refers
to local fluctuations in the graph topology, such as the number of neighbors coupled
to each spin (the so-called degrees \cite{Newman2001}).

Universal scaling around phase transitions is perhaps the most striking collective property of spin models \cite{CardyBook}. Renormalization group theory and
the computation of critical exponents have prompted the
notion of  universality classes, i.e., the fact that systems very different in nature share the same critical behavior.
Understanding how network heterogeneities modify the critical properties of spin models is a central problem in network science \cite{Doro2008}.
Specifically, the equilibrium critical behavior of the Ising model on networks is characterized by mean-field critical
exponents \cite{Leone2002,Doro2002,Dommers2016}, as long as the fourth moment of the degree distribution is finite.

Progress has been much slower on the side of the dynamical critical properties of spin models on networks.
To our knowledge, the effect of network heterogeneities on the dynamical exponents \cite{Odor2004} of the Ising model is not known.
More than that, depending on the details of the dynamics and of the network structure \cite{Coolen2001Stat,Coolen2001Dyn,Torrisi2022}, the Ising model
may evolve to nonequilibrium stationary states 
that do not follow the Boltzmann distribution. In this case, even if the interest
lies only on the stationary critical properties, one has to abandon equilibrium statistical mechanics and shift to a full dynamical approach.

The nonequilibrium dynamics of the Ising model on random graphs has been exactly solved in the thermodynamic limit using the generating
functional approach \cite{Hatchett2004,Mimura2009} and the cavity method \cite{Neri2009}.
Even though the microscopic dynamics of the model is by construction a Markovian process, symmetric couplings among the spins
induce retarded self-interactions and the formal solution of the problem 
is a path-probability for the effective dynamics
of a single spin \cite{Neri2009}.
The history dependency encoded in the path-probability prevents any attempt to calculate analytically
the trace over the single-spin configurations.
Besides that, the dimension of the path-probability  grows exponentially in time, which quickly renders numerical computations unfeasible. These features make
the dynamics of spin models on networks a notorious difficult problem, which has 
stimulated the design of various approximate methods.
Some of them rely on assumptions to reduce the number of variables in the problem
and obtain a closed set of dynamical equations  \cite{Mozeika2008,Metz2008,Ferraro2015,Aurell2017}, while other approaches, such as
the dynamical TAP equations \cite{Roudi2011,Roudi2011a,Aurell2012}
and cluster variational methods \cite{Pelizzola2013,Vazquez2017}, are inspired in well established methods for the equilibrium properties of spin models.

There are two main classes of graphs for which
the effective problem simplifies and one can derive closed-form dynamical equations:
dense random graphs and
sparse directed random graphs  \cite{Derrida1987,Neri2009,Mimura2009}. In the former case, each spin is densely connected with
the rest of the network and the path probability simplifies on account of the 
law of large numbers.
In the second case, the absence of bidirected edges eliminate the history dependency and the effective dynamics becomes Markovian.
In both cases, the exact dynamics follows from a simplified form of the cavity equations for the path probability \cite{Neri2009}.

The cavity or message-passing equations provide an algorithm to compute the local marginals of a variety of problems
defined on random graphs \cite{Mezard2001,Mezard2003,Mezard2009}.
In general, these equations do not admit analytic solutions on undirected graphs with an heterogeneous structure.
However, the spectra of undirected random graphs \cite{Metz2020,Silva2022} and the equilibrium of spin models on networks \cite{Metz2022} have
been recently studied by means
of an interesting family of analytic solutions of the cavity equations, in which the mean degree
is infinitely large, but the solutions still depend on the full degree distribution.
This class of solutions is simple enough that it allows to address the role of degree fluctuations
in a comprehensive way. Since the cavity equations share the same formal structure across different areas \cite{Mezard2009}, one expects
to extract an analogous solution for the dynamics of the Ising model on heterogeneous networks.

Here we confirm this expectation and derive an exact solution for the synchronous dynamics of the Ising model on
highly-connected random graphs with an arbitrary degree distribution. 
The stochastic dynamics of the spins is governed by an arbitrary distribution of the threshold
noise that mimics the contact of the system with a thermal bath. Depending on the choice of the distribution
of thresholds \cite{Coolen2001Stat}, the model evolves to stationary states that are not described by the Boltzmann distribution.
Therefore, the present model allows to clearly study how degree fluctuations and the nonequilibrium nature of the stationary states
influence the critical properties of the Ising model.
Besides that, networks of binary units with random thresholds give valuable
insights into neural networks \cite{Coolen2001Stat,Coolen2001Dyn}, choice and opinion dynamics \cite{Castellano2009,Kanoria2011,Holehouse2022}, gene regulatory
networks \cite{Kauffman1969,Mozeika2011,Torrisi2022,Hurry2022}, and socio-economic phenomena \cite{Bouchaud2013}.

We derive a simple dynamical equation for the full distribution of local magnetizations, from which we find the critical line that
separates the paramagnetic from the ferromagnetic region.
We compute stationary and dynamical critical exponents
for the mean and the variance of local magnetizations in the case of a negative binomial
degree distribution \cite{Metz2022,Silva2022}.
For an hyperbolic tangent distribution of thresholds, for which the stationary states follow a Boltzmann-like
distribution \cite{Peretto1984}, all critical indexes assume their standard mean-field values \cite{Janssen1989,Odor2004,Calabrese2005,Adzhemyan2022}.
In contrast, when the threshold noise follows an algebraic distribution and detailed
balance is presumably broken, the stationary
critical behavior and the long-time critical dynamics are both characterized by the same values of the critical indexes, which
are determined by the power-law tails of the distribution of thresholds.
On the other hand, the characteristic time for the exponential relaxation
of the average magnetization inside each phase always exhibits a mean-field critical behavior \cite{Odor2004}, independently of the distribution of thresholds.
Lastly, we derive analytic expressions for the stationary distribution of local magnetizations
inside the ferromagnetic phase and we show that its variance displays a maximum as a function
of the temperature, due to the interplay between threshold noise and degree fluctuations. Some of our theoretical findings
are corroborated by numerical simulations.

The paper is organized as follows. In the next section we define the model and its microscopic dynamics. Section \ref{secder} explains
how to obtain the recurrence equation for the dynamics of the distribution of local magnetizations. We present
the results for the critical exponents in section \ref{sec:results}, and some final remarks in section \ref{sec:final}. The paper
contains an appendix that explains how to solve the dynamics using the generating functional approach.

%%%%%%%%%%%%%%%%%%%%%%%%%%%%%%%%%%%%%%%%%%%%%%%%%%%%%%%%%%%%%%%%%%%%%%%%%%%%%%%%%%%%%%%%%%%%%%%%%%
%%%%%%%%%%%%%%%%%%%%%%%%%%%%%%%%%%%%%%%%%%%%%%%%%%%%%%%%%%%%%%%%%%%%%%%%%%%%%%%%%%%%%%%%%%%%%%%%%%%
%%%%%%%%%%%%%%%%%%%%%%%%%%%%%%%%%%%%%%%%%%%%%%%%%%%%%%%%%%%%%%%%%%%%%%%%%%%%%%%%%%%%%%%%%%%%%%%%%%

\section{Model definitions}
\label{sec:model}

We study the dynamics of $N$ Ising spins $\sigma_{i}(t) \in \{ -1,1 \}$ ($i=1,\dots,N$) that interact through the edges of an undirected and simple random graph \cite{Bollo2001}. The states
evolve in time $t$ by following a Markov process, in which $t=0,1,\dots,t_{\rm max}$ is a discrete variable and all spins are synchronously updated according
to their local fields at the previous time step
\begin{equation}
  \sigma_{i}(t+1) = {\rm sign} \left( h_i \left[ \boldsymbol{\sigma}(t) \right] + T \zeta_{i}(t)  \right) ,
  \label{jsa}
\end{equation} 
where $\{ \zeta_{i}(t) \}$ are independent and
identically distributed random variables drawn from a distribution $\mu(\zeta)$ that fulfills $\mu(-\zeta) =\mu(\zeta)$ . The temperature $T \geq 0$ controls
the threshold noise in the stochastic dynamics: for $T=0$ the dynamics is deterministic, whereas for $T \rightarrow \infty$ it is completely random.

The local field $h_i[\boldsymbol{\sigma}(t)]$ at time $t$ due to the global state $\boldsymbol{\sigma}(t) = (\sigma_{1}(t),\dots,\sigma_{N}(t))$ is given by
\begin{equation}
  h_i[\boldsymbol{\sigma}(t)] = \frac{J}{c} \sum_{j=1}^N C_{ij} \sigma_{j}(t),
  \label{loc}
\end{equation}  
where the binary random variables $C_{ij} \in \{0,1\}$ are the elements of the adjacency matrix $\mathbf{C}$ that specifies the topology of
the random graph model. If there
is an interaction between the spins located at nodes $i$ and $j$, then we set $C_{ij}=1$, whereas $C_{ij}=0$ if the corresponding spins do not interact. The matrix
$\mathbf{C}$ is symmetric (the graph is undirected) and its diagonal entries are zero. The parameter $J > 0$ denotes the strength of the pairwise ferromagnetic
interactions between adjacent spins, while $c$ is the so-called mean degree (see below) or average coordination number. The scaling of the coupling strengths with $c$ is suitable
to analyze the model in the limit $c \rightarrow \infty$.

In order to derive a discrete map for the time evolution of the global magnetization,
\begin{equation}
m(t) =  \frac{1}{N} \sum_{i=1}^N \sigma_{i}(t),
\end{equation}  
it is more convenient to formulate the dynamics in terms of probabilities. Equation (\ref{jsa}) defines a Markov process and the probability
$p(\boldsymbol{\sigma},t)$ to observe a global configuration $\boldsymbol{\sigma} = (\sigma_1,\dots,\sigma_N)$ at time $t$ evolves as follows
\begin{equation}
  p(\boldsymbol{\sigma},t+1) = \sum_{\boldsymbol{\sigma}^{\prime}} W(\boldsymbol{\sigma}|\boldsymbol{\sigma}^{\prime}) p(\boldsymbol{\sigma}^{\prime},t),
  \label{prob1}
\end{equation}  
where$\sum_{\boldsymbol{\sigma}^{\prime}}$ runs over the $2^N$ configurations of the system, and the matrix element
$W(\boldsymbol{\sigma}|\boldsymbol{\sigma}^{\prime})$ is the conditioned probability to observe a transition
from state $\boldsymbol{\sigma}^{\prime}$ to $\boldsymbol{\sigma}$.
By integrating over $\zeta_{i}(t)$ in Eq. (\ref{jsa}), one finds the explicit form
of the transition matrix elements $W(\boldsymbol{\sigma}|\boldsymbol{\sigma}^{\prime})$
\begin{equation}
\label{transitions}
W(\boldsymbol{\sigma}|\boldsymbol{\sigma}^{\prime}) = \prod_{i=1}^N \frac{1}{2} \left( 1 + \sigma_i  \mathcal{F} \left[ \beta h_i(\boldsymbol{\sigma}^{\prime}) \right]  \right),
\end{equation}  
where $\beta = T^{-1}$, and $\mathcal{F}(x)$ is determined by the distribution $\mu(\zeta)$ of the threshold noise as follows
\begin{equation}
\mathcal{F}(x) = \int_{-x}^{x} d \zeta  \, \mu(\zeta).
\end{equation}  
The function $\mathcal{F} (x)$ satisfies the properties 
\begin{equation}
  \mathcal{F}(-x) = - \mathcal{F}(x), \qquad \lim_{x \rightarrow \pm \infty} \mathcal{F}(x) = \pm 1.
  \label{ggpp}
\end{equation}  
Depending on the choice of $\mu(\zeta)$, the stationary distribution
of the spin configurations is not given by the Boltzmann distribution and we
expect that detailed balance breaks down \cite{Coolen2001Stat}.

Let us now  specify the random graph ensemble.
The coordination number or degree $K_i$ of node $i$, defined in terms
of $\mathbf{C}$ as
\begin{equation}
K_i = \sum_{j=1}^N C_{ij},
\end{equation}  
gives the number of spins coupled to $\sigma_i$.
One of our purposes is to investigate how fluctuations in the degree sequence $K_1,\dots,K_N$ impact
the stationary and the dynamical critical properties of the Ising model.
Thus, we consider the configuration model of networks \cite{Molloy1995,Newman2001,Fosdick_2018}, in which the
degrees $K_1,\dots,K_N$ are independently drawn from a common distribution
\begin{equation}
p_k = \lim_{N \rightarrow \infty} \frac{1}{N} \sum_{i=1}^N \delta_{K_i,k},
\end{equation}  
and a single graph instance is generated by randomly choosing pairs of nodes
and then connecting them subject to the prescribed degrees.
The first moment of $p_k$ yields the average degree 
\begin{equation}
c = \sum_{k=0}^{\infty} k p_k,
\end{equation}  
which provides the mean number of neighbors coupled to a single spin.
Since the degree distribution $p_k$ is specified at the outset, the configuration model provides
the ideal setting to explore the effect of degree fluctuations by changing the shape of $p_k$. 

In the next section, we exactly solve the nonequilibrium dynamics of this model in the limit $N \rightarrow \infty$ for arbitrary distributions $p_k$ and $\mu(\zeta)$. The solution is valid
in the high-connectivity limit $c \rightarrow \infty$, provided $c/N \rightarrow 0$. This regime is achieved by first taking the limit $N \rightarrow \infty$
and then $c \rightarrow \infty$ afterwards \cite{Metz2022}.

%%%%%%%%%%%%%%%%%%%%%%%%%%%%%%%%%%%%%%%%%%%%%%%%%%%%%%%%%%%%%%%%%%%%%%%%%%%%%%%%%%%%%%%%%%
%%%%%%%%%%%%%%%%%%%%%%%%%%%%%%%%%%%%%%%%%%%%%%%%%%%%%%%%%%%%%%%%%%%%%%%%%%%%%%%%%%%%%%%%%%%%

\section{Recurrence equations for the dynamics}
\label{secder}

In this section we derive an exact map for the time evolution of the global magnetization and for the full distribution of local magnetizations in the limit $c \rightarrow \infty$ by
using the law of large numbers.
As a byproduct, we put forward an effective approximation for the interactions between the spins
valid for large $c$. In the appendix \ref{app1}, we present a more rigorous
derivation of the dynamical equation for the global magnetization by using the generating functional approach \cite{Martin1973,Coolen2001Dyn}.

The probability $p_{i}(\sigma,t)$ of observing the spin at site $i$ in the state $\sigma \in \{ -1, 1 \}$  at time $t$ follows
from the marginalization
\begin{equation}
  p_{i}(\sigma,t) = \sum_{\boldsymbol{\sigma} \setminus \sigma_i} p(\boldsymbol{\sigma},t),
  \label{prob122}
\end{equation}  
where $\sum_{\boldsymbol{\sigma} \setminus \sigma_i}$ sums over the configurations of all spins except for $\sigma_i$.  The  local magnetization $m_{i}(t)$ at time $t$ reads
\begin{equation}
m_{i}(t) = \sum_{\sigma} \sigma   p_{i}(\sigma,t) ,
\end{equation}  
while the global magnetization $m(t)$ is given by
\begin{equation}
m(t) = \frac{1}{N} \sum_{i=1}^N m_{i}(t).
\end{equation}  
Our primary aim is to obtain an exact recursive equation for $m(t)$ in the thermodynamic limit $N \rightarrow \infty$. Inserting Eq. (\ref{prob1}) in Eq. (\ref{prob122}) and using
the explicit form of $W(\boldsymbol{\sigma}|\boldsymbol{\sigma}^{\prime})$, Eq. (\ref{transitions}), one
can write the local magnetization as
\begin{equation}
  m_{i}(t+1) =      \sum_{\boldsymbol{\sigma}} p(\boldsymbol{\sigma},t)  \mathcal{F}\left( \frac{\beta J}{ c} \sum_{j \in \partial_i} \sigma_j   \right),
  \label{hdka}
\end{equation}  
where $\partial_i$ represents the set of nodes that are adjacent to node $i$.
Since the sum over $j \in \partial_i$ contains a number of terms of order $\mathcal{O}(K_i)$,  we invoke the law of large
numbers for $c \rightarrow \infty$  and replace this sum by the expectation value
\begin{equation}
  \frac{1}{K_i} \sum_{j \in \partial_i} \sigma_j \xrightarrow{c \rightarrow \infty}   u(t)  =  \sum_{\sigma} P(\sigma,t) \sigma,
  \label{gaqw}
\end{equation}  
where 
\begin{equation}
  P(\sigma,t) = \frac{\sum_{ij=1}^N C_{ij} \sum_{\sigma^{\prime}} p_{i}(\sigma^{\prime},t) \delta_{\sigma^{\prime} \sigma }}{ \sum_{ij=1}^N C_{ij}  }
  \label{gdq1}
\end{equation}  
is the probability that a randomly chosen edge has one of its spins in the state $\sigma$ at time $t$. Thus, the spatial fluctuations
of the local field on the right hand side of Eq. (\ref{hdka}) are solely determined by the degree distribution. In the limit $c \rightarrow \infty$, the local
magnetization fulfills
\begin{equation}
  m_{i}(t+1) =  \mathcal{F} \left[\beta J G_i u(t)  \right],
  \label{huhu}
\end{equation}  
where the rescaled degrees $G_i = K_i/c$ ($i=1,\dots,N$) are distributed as follows
\begin{equation}
  \nu(g) = \lim_{c \rightarrow \infty} \sum_{k=0}^{\infty} p_k  \delta\left(g - \frac{k}{c}   \right).
  \label{nu}
\end{equation}  
In terms of $\nu(g)$, the global magnetization at time $t+1$ is determined by recurrence equation
\begin{equation}
  m(t+1) = \int_{0}^{\infty} d g \,  \nu(g)  \, \mathcal{F} \left[ \beta J  g u(t)  \right].
  \label{m1}
\end{equation}  
Equations (\ref{huhu}) and (\ref{m1}) are valid when the variance of the rescaled degrees remains finite in the high-connectivity limit. One can access this connectivity
regime for $N \rightarrow \infty$ by setting  $c \propto N^{a}$ ($0 < a < 1$) \cite{Metz2022}. In contrast, one recovers
the dynamics of a fully-connected system in the dense 
regime $c \propto N$ ($N \rightarrow \infty$), for which degree fluctuations are irrelevant.

In order to determine $m(t+1)$, we need to find the recurrence equation for $u(t)$. Inserting Eq. (\ref{gdq1}) in the definition
of $u(t)$, we obtain
\begin{equation}
  u(t+1) = \frac{\sum_{ij=1}^N  C_{ij} m_{i}(t+1) }{\sum_{ij=1}^N  C_{ij} }.
  \label{bua1}
\end{equation}  
For large $c$, we can use Eq. (\ref{huhu}) and rewrite $u(t+1)$ as follows
\begin{equation}
  u(t+1) = \frac{\sum_{i=1}^N K_i  \mathcal{F} \left[ \beta J G_i u(t)  \right] }{\sum_{i=1}^N  K_i }.
  \label{hh5}
\end{equation}  
We see that $u(t)$ is a global observable that weights the local magnetization of each site according to its rescaled degree $G_i = K_i/c$. In the
limit $c \rightarrow \infty$, the above equation is rewritten as
\begin{equation}
  u(t+1) = \int_{0}^{\infty} d g \, \nu(g) \,  g \,  \mathcal{F} \left[ \beta J  g u(t)  \right].
   \label{m2}
\end{equation}
Equations (\ref{m1}) and (\ref{m2}) describe the nonequilibrium dynamics of the global magnetization of the Ising model on infinitely large random
graphs in the high-connectivity limit $c \rightarrow \infty$. This solution is valid for arbitrary distributions $\nu(g)$ and $\mu(\zeta)$. In the appendix \ref{app1}, we present
a formal derivation of such equations by using the generating functional approach pioneered  in \cite{Martin1973}.

In contrast to fully-connected models, the local magnetizations on heterogeneous random graphs fluctuate from
site to site. Equation (\ref{huhu}) fully characterizes the spatial fluctuations
of $m_i(t)$ at any time step $t$. The probability distribution $\mathcal{P}(m,t)$
of the local magnetization at time $t$ is determined only by $\nu(g)$ and by the activation function $\mathcal{F}(x)$. By making a simple change
of variables, we find a formal recurrence relation for $\mathcal{P}(m,t)$
\begin{equation}
  \mathcal{P}(m,t+1) = \frac{1}{\beta J u(t)} \frac{d \mathcal{F}^{-1}(m)}{d m} \nu \left[ \frac{\mathcal{F}^{-1}(m) }{\beta J u(t) }    \right],
  \label{huhu1}
\end{equation}  
in which $\mathcal{F}^{-1}(x)$ is the inverse of $\mathcal{F}(x)$ under composition. The above equation allows to study the nonequilibrium dynamics of the distribution $\mathcal{P}(m,t)$.

Interestingly, we can also compute the stationary distribution
\begin{equation}
p_{\infty}(\boldsymbol{\sigma}) \equiv \lim_{t \rightarrow \infty}   p(\boldsymbol{\sigma},t)
\end{equation}
of the spin configurations in the high-connectivity limit. The object $p_{\infty}(\boldsymbol{\sigma})$
fulfills the self-consistent equation
\begin{equation}
  p_{\infty}(\boldsymbol{\sigma}) = \sum_{\boldsymbol{\sigma}^{\prime}} W(\boldsymbol{\sigma}|\boldsymbol{\sigma}^{\prime}) p_{\infty}(\boldsymbol{\sigma}^{\prime}).
  \label{hsdp}
\end{equation}
Thus, in the stationary regime, $\boldsymbol{\sigma}^{\prime}$ in the local field $h_{i}(\boldsymbol{\sigma}^\prime)$ of Eq. (\ref{transitions}) is sampled
from  $p_{\infty}(\boldsymbol{\sigma}^{\prime})$. By using the law of large numbers, we can write
\begin{equation}
  h_{i}(\boldsymbol{\sigma}^\prime) = \frac{J}{c} \sum_{j \in \partial_i} \sigma_{j}^{\prime} \xrightarrow{c \rightarrow \infty}  J G_i u,
  \label{hdga22}
\end{equation}  
where we have assumed that $u(t)$ evolves to a fixed-point $u$ when $t \rightarrow \infty$.
From Eqs. (\ref{transitions}) and (\ref{hdga22}), we conclude that $W(\boldsymbol{\sigma}|\boldsymbol{\sigma}^{\prime})$ becomes independent of
$\boldsymbol{\sigma}^{\prime}$ for large $c$, which immediately
leads to 
\begin{equation}
  p_{\infty}(\boldsymbol{\sigma}) = \prod_{i=1}^N \frac{1}{2} \left[ 1 + \sigma_i \mathcal{F} \left( \beta J G_i u \right)  \right].
  \label{hdg}
\end{equation}  
The above equation describes the stationary distribution of the spins for a single realization of the graph
in which both $c$ and $N$ are very large, but the ratio $c/N$ is vanishing small.

Equation (\ref{hdg}) explicitly depends on the distribution
$\mu$ of the threshold noise by means of $\mathcal{F}(x)$.
For the particular choice $\mathcal{F}(x)=\tanh(x)$, the
long-time synchronous dynamics fulfills detailed balance and Eq. (\ref{hdg}) corresponds to a Boltzmann-like
distribution \cite{Peretto1984,Coolen2001Stat}, which enables the application of equilibrium statistical mechanics.
Indeed, by starting from the standard form of the equilibrium
distribution $p_{\infty}(\boldsymbol{\sigma}) \sim \prod_{i=1} \cosh{\left[\beta h_{i}(\boldsymbol{\sigma})  \right]}$ for synchronous
dynamics \cite{Coolen2001Stat}, one can duplicate the configuration space, apply the law of large numbers on the local fields in the high-connectivity limit, and then recover Eq. (\ref{hdg}) when
$\mathcal{F}(x)=\tanh(x)$.
The fact that Eq. (\ref{hdg}) is not generally given by a Boltzmann-like form strongly indicates
that detailed balance breaks down depending on the choice of $\mathcal{F}(x)$ \cite{Coolen2001Stat}.

We end this section by presenting an useful approximation for the interaction matrix of the Ising model on random graphs.
We note from Eqs. (\ref{huhu}) and (\ref{hh5})  that the local field at node $i$ can be written for large $c$ as
\begin{equation}
  h_i \left[\boldsymbol{\sigma}(t) \right]  = \frac{J}{N}  \sum_{j=1}^N G_i G_j m_j(t).
  \label{potu1}
\end{equation}  
It follows from Eqs. (\ref{loc}) and (\ref{potu1}) that the entries
of the original adjacency matrix $\mathbf{C}$ can be replaced, for large enough $c$, by the effective matrix elements
\begin{equation}
  C_{ij}^{{\rm eff}} = \frac{c}{N} G_i G_j (i \neq j).
  \label{kda1}
\end{equation}  
The above equation provides a practical way to simulate the adjacency matrix of the ferromagnetic Ising model
on highly-connected random graphs with arbitrary degree distributions, without
having to generate random graph instances using more sophisticated algorithms \cite{Fosdick_2018,CoolenBookGraphs}.
Equation (\ref{kda1}) defines a complete graph that is essentially equivalent to the graph ensemble studied in \cite{Dommers2016}.
Below we confirm the exactness of Eq. (\ref{kda1}) by comparing
our theoretical recurrence equation for the average magnetization with numerical simulations.

%%%%%%%%%%%%%%%%%%%%%%%%%%%%%%%%%%%%%%%%%%%%%%%%%%%%%%%%%%%%%%%%%%%%%%%%%%%%%%%%%%%%%%%%%%%%%%%%%%%%%
%%%%%%%%%%%%%%%%%%%%%%%%%%%%%%%%%%%%%%%%%%%%%%%%%%%%%%%%%%%%%%%%%%%%%%%%%%%%%%%%%%%%%%%%%%%%%%%%%%

\section{Results}
\label{sec:results}

Equations (\ref{m2}) and (\ref{huhu1}) describe the dynamics of the Ising model on an infinitely large random graph
with an arbitrary degree distribution $\nu(g)$ and for any symmetric distribution $\mu (\zeta)$ of the threshold noise.
In this section we present results for the stationary and the dynamical critical properties of the model in the
case of a negative binomial degree distribution
\begin{equation}
  p_{k}^{\rm bin} = \frac{\Gamma(\alpha + k )   }{k! \Gamma(\alpha)}  \left(  \frac{c}{\alpha} \right)^k  \left(  \frac{\alpha}{\alpha + c} \right)^{\alpha + k},
  \label{tuods}
\end{equation}  
which is parametrized by $0 < \alpha < \infty$ and the mean degree $c$.
The variance $\sigma^2$ of $p_{k}^{\rm bin}$ is given by
\begin{equation}
\sigma^2 = c + \frac{c^2 }{\alpha}.
\end{equation}  
We recover the geometric degree distribution \cite{Newman2001} and the Poisson degree distribution by setting, respectively, $\alpha=1$ and $\alpha \rightarrow \infty$.
Given that
\begin{equation}
   \lim_{c \rightarrow \infty}  \frac{\sigma^2}{c^2} = \frac{1}{\alpha},
  \label{js33}
\end{equation}  
the relative variance of $p_{k}^{\rm bin}$ is controlled only by $\alpha$ in the high-connectivity limit, which renders the negative binomial
distribution very convenient to probe the effect of heterogeneous degrees on the dynamics. Substituting Eq. (\ref{tuods}) in Eq. (\ref{nu}), we
can find the explicit form of $\nu(g)$
\begin{equation}
\nu_{\rm bin}(g) = \frac{\alpha^{\alpha}}{\Gamma(\alpha)} g^{\alpha-1} e^{- \alpha g}.
\end{equation}  
In the limit $\alpha \rightarrow \infty$, the relative variance goes to zero 
and we expect to recover the recurrence equations for the dynamics
of the Curie-Weiss model \cite{Coolen2001Dyn}.

We will present numerical results for two different distributions of the threshold noise
\begin{eqnarray}
&\mu_{\rm h}(\zeta)& = \frac{1}{2} \left[ 1 - \tanh^{2} \left( \zeta \right)  \right],  \label{utq1} \\
&\mu_{\rm a}(\zeta)& = \frac{1}{2 \left( 1 + \zeta^{2 \kappa}  \right)^{1 + \frac{1}{2 \kappa } }  } , \label{utq2}
\end{eqnarray}  
where $\kappa$ in $\mu_{\rm a}(\zeta)$ is a positive integer which ensures that the symmetry $\mu_{\rm a}(\zeta) = \mu_{\rm a}(-\zeta)$ is
preserved. More precisely, we will discuss results for $\kappa \in \{ 1,2,3 \}$.
The corresponding activation functions are given by
\begin{eqnarray}
  \mathcal{F}_{\rm h} (x) &=& \tanh{(x)}, \label{poq1} \\
  \mathcal{F}_{\rm a} (x)  &=& \frac{x}{\left( 1 + x^{2 \kappa}  \right)^{\frac{1}{2 \kappa}}}.   \label{poq2} 
\end{eqnarray}  
The hyperbolic tangent distribution $\mu_{\rm h}(\zeta)$ has exponential tails, while the algebraic distribution $\mu_{\rm a}(\zeta)$ behaves
as $\mu_{\rm a}(\zeta) \propto |\zeta|^{-2 \kappa - 1}$ for $|\zeta| \gg 1$. Consequently, the $n$-th moment of $\mu_{\rm a}(\zeta)$ diverges 
if $n \geq 2 \kappa$. In addition, if the threshold noise is sampled from $\mu_{\rm h}$, 
the stationary spin configurations follow
a Boltzmann-like distribution, obtained from Eqs. (\ref{hdg}) and (\ref{poq1}).
In contrast, the long-time behavior of the system is not described by a Boltzmann-like distribution if the threshold noise
follows from $\mu_{\rm a}$. In this case the system reaches nonequilibrium stationary states and equilibrium
statistical mechanics is not applicable.
Thus, our choices of $\mu$ allow us to examine how strong fluctuations in the threshold noise and the concomitant 
absence of Boltzmann equilibrium impact the dynamics
and the stationary states of the Ising model.

%%%%%%%%%%%%%%%%%%%%%%%%%%%%%%%%%%%%%%%%%%%%%%%%%%%%%%%%%%%%%%%%%%%%%%%%%%%

\subsection{Stationary behavior}

The fixed-point equations of the dynamics are obtained by setting $\lim_{t \rightarrow \infty} u(t) = u$ and $\lim_{t \rightarrow \infty} m(t) = m$ in Eqs. (\ref{m1}) and (\ref{m2})
\begin{eqnarray}
  m = \int_{0}^{\infty} d g \, \nu(g) \,  \mathcal{F} \left( \beta J g u  \right), \label{ja1} \\
u = \int_{0}^{\infty} d g \, \nu(g) \,  g \,  \mathcal{F} \left( \beta J  g u  \right).\label{ja2}
\end{eqnarray}  
The above expressions generalize the standard mean-field description of the Curie-Weiss model \cite{Kochma2013}.
By setting  $\mathcal{F}(x) = \mathcal{F}_{\rm h}(x)$, we recover the fixed-point equations derived through
equilibrium statistical mechanics in reference \cite{Metz2022}. 

Since $\mathcal{F}(0)=0$, Eqs. (\ref{ja1}) and (\ref{ja2}) admit a paramagnetic solution $u = m =0$. By expanding the
integrand of Eq. (\ref{ja2}) up to $O(u)$, we find that a nontrivial solution $|u| > 0$ appears below the critical temperature
\begin{equation}
  T_c = J A_{\mu} (1 + \Delta_{\nu}^2),
  \label{tutu}
\end{equation} 
where
\begin{equation}
\Delta_{\nu}^2 = \int_{0}^{\infty} d g g^2 \nu(g) - 1, \qquad A_{\mu}= \frac{d \mathcal{F}}{d x}\Big{|}_{x=0} .
\end{equation}  
Equation (\ref{tutu}) is valid for arbitrary distributions $\nu(g)$ and $\mu(\zeta)$, and it
shows that the model has a finite critical temperature when the variance of $\nu(g)$ is finite \cite{Leone2002,Doro2002,Dommers2016}.
The tails of $\mu(\zeta)$ are irrelevant for the critical temperature $T_c$, which depends on $\mu(\zeta)$ only through its behavior around $\zeta=0$.
For $\Delta_{\nu}^2 = 0$ and $A_{\mu} = 1$, we obtain the critical temperature $T_c = J$ of
the Curie-Weiss model \cite{Kochma2013}. Regular random graphs and Erd\"os-R\'enyi random graphs \cite{Bollo2001} are the most representative
homogeneous random graph ensembles for which $\Delta_{\nu}^2 = 0$.

The system undergoes a continuous transition between the ferromagnetic $(|m| >0)$ and the
paramagnetic $(m=0)$ phase at $T=T_c$. The stationary order-parameter $u$ determines the global magnetization $m$ and all
moments of the stationary distribution $\mathcal{P}(m) = \lim_{t \rightarrow \infty} \mathcal{P}(m,t)$ of
local magnetizations. For $u > 0$ ($u < 0$) the support of $\mathcal{P}(m)$ is the interval $m \in [0,1]$ ($m \in [-1,0]$).
By setting $|u| \neq 0$, we obtain from Eqs. (\ref{huhu1}), (\ref{poq1}), and (\ref{poq2}) the
corresponding distributions 
\begin{equation}
  \mathcal{P}_{\rm h}(m) = \frac{1}{\beta J |u| (1-m^2)} \nu \left[ \frac{\tanh^{-1}(m)}{\beta J u }   \right]
  \label{hua1}
\end{equation}  
and
\begin{equation}
  \mathcal{P}_{\rm a}(m) = \frac{1}{\beta J |u| (1-m^{2 \kappa})^{1 + \frac{1}{2 \kappa} } } \nu \left[ \frac{m}{\beta J u \left( 1 - m^{2 \kappa}   \right)^{\frac{1}{2 \kappa }} }   \right]
  \label{hua2}
\end{equation}  
for the hyperbolic and the algebraic activation function, respectively.
For homogeneous random graphs, in which $\nu(g) = \delta(g-1)$, the above
equations yield the $\delta$-distribution $\mathcal{P}(m) = \delta \left(m - \mathcal{F}(\beta J m)  \right)$.

\begin{figure}[htp]
  \begin{center}
    \includegraphics[scale=0.55]{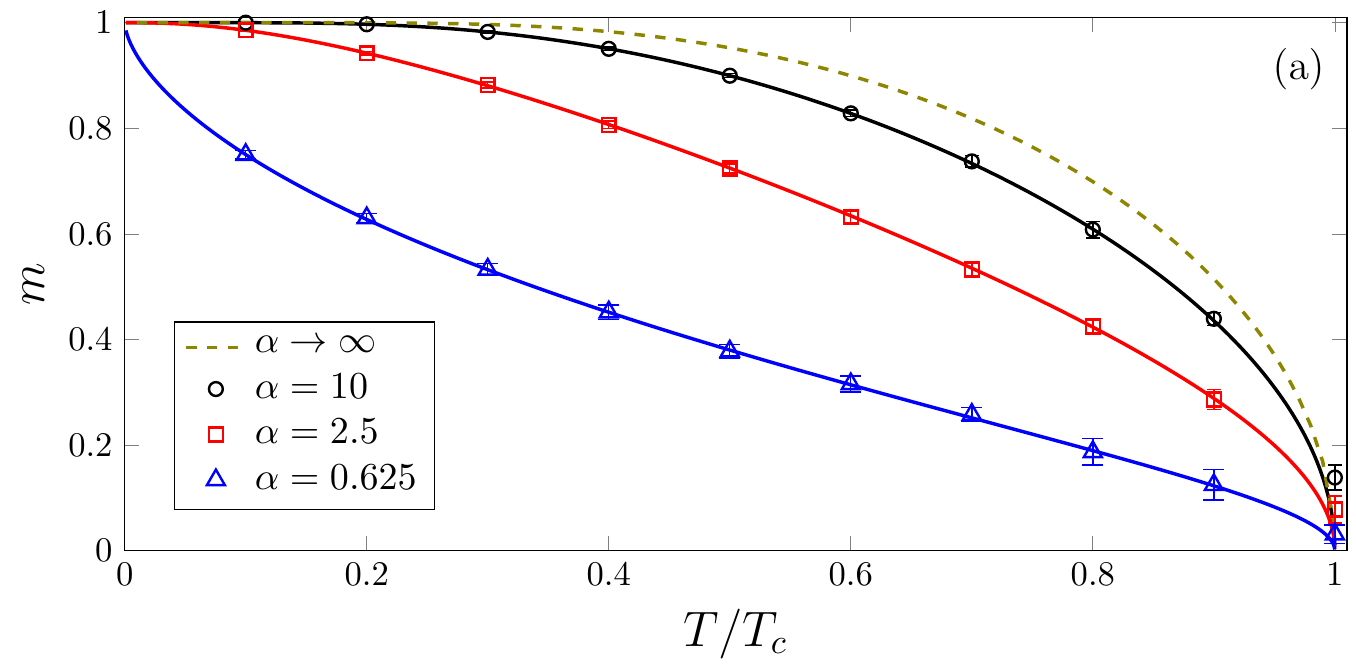} \\
    \hspace{-0.4cm}
    \includegraphics[scale=0.55]{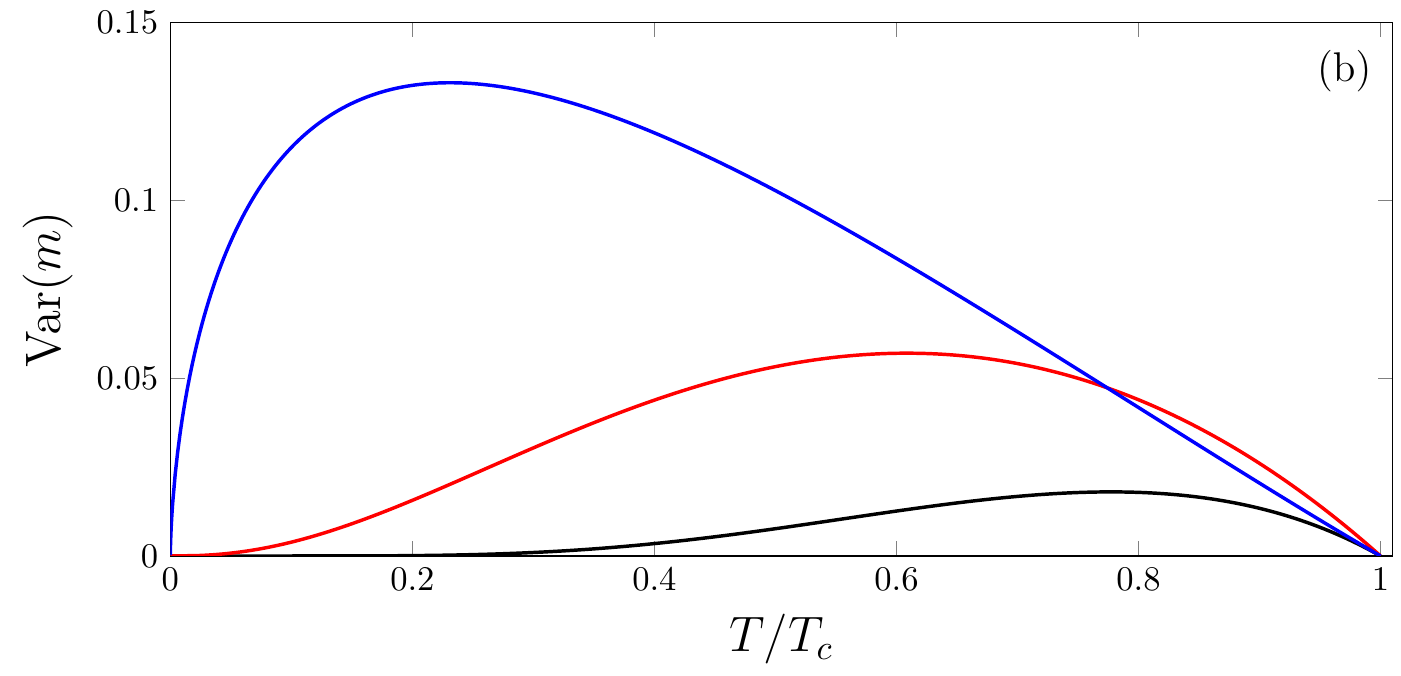}
    \caption{(a) Stationary magnetization $m$ and (b) variance ${\rm Var}(m)$ of the distribution $\mathcal{P}_{\rm h}(m)$ of local magnetizations as a function of the
      rescaled temperature $T/T_c$ (see Eq. (\ref{tutu})) for the hyperbolic tangent distribution $\mu_{\rm h}$ of the threshold noise.
      The parameter $1/ \alpha$ is the relative variance of the negative binomial degree distribution (see Eq. (\ref{js33})). The solid lines follow from the analytic
      Eqs. (\ref{ja1}), (\ref{ja2}), and (\ref{hua1}). The symbols are results obtained from numerical simulations of Eq. (\ref{jsa}) for
      $N=10^{4}$ and mean degree $c=10^2$. The vertical bars are the standard deviation calculated from
      $10$ independent runs of the simulations. The random graph samples in the
      numerical simulations are generated from Eq. (\ref{kda1}).
    }
\label{mandT}
\end{center}
\end{figure}

\begin{figure}[htp]
  \begin{center}
    \includegraphics[scale=0.58]{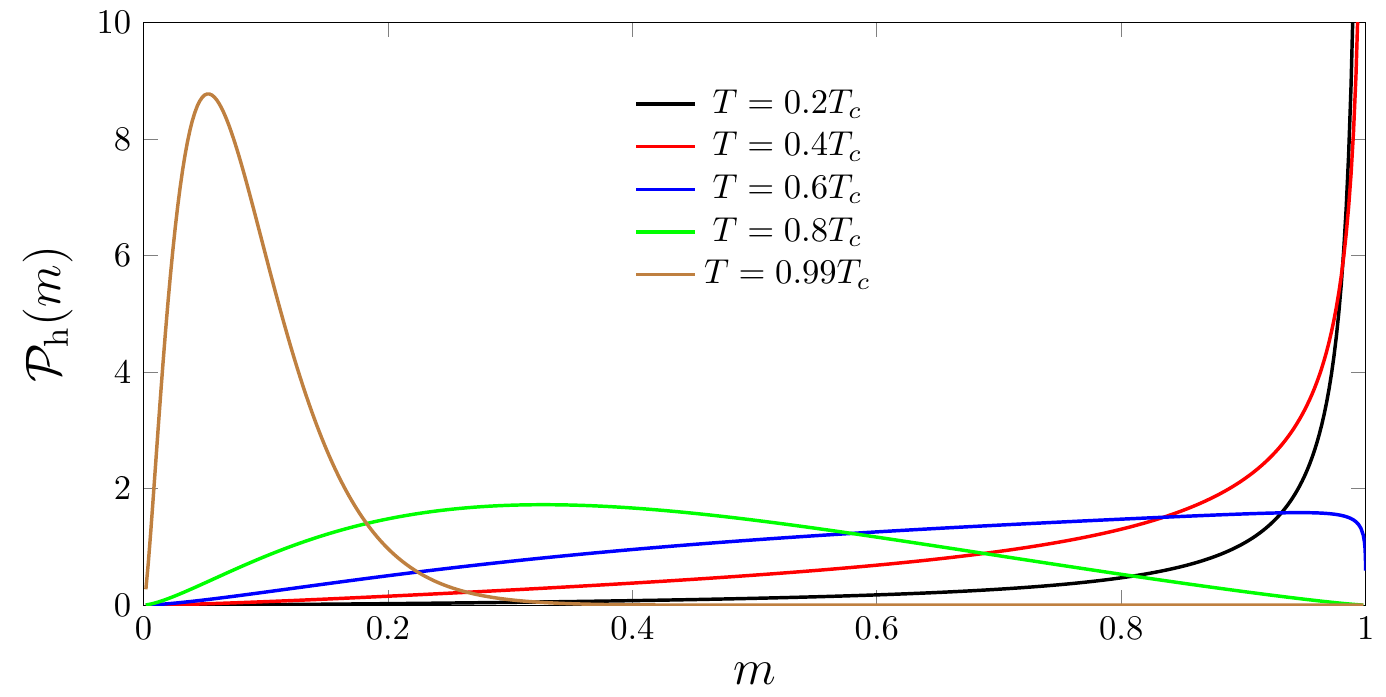} 
    \caption{The stationary distribution $\mathcal{P}_{\rm h}(m)$ of local magnetizations inside the ferromagnetic phase of the Ising model on random graphs
      with a negative binomial degree distribution with $\alpha=2.5$ (see Eq. (\ref{js33})) and the hyperbolic
      tangent distribution $\mu_{\rm h}$ of the threshold noise.
    }
\label{distrP}
\end{center}
\end{figure}

Figure \ref{mandT} shows the global magnetization
$m$ and the variance ${\rm Var}(m)$ of $\mathcal{P}(m)$ as a function of the temperature $T$ for the hyperbolic distribution of the threshold noise and
different values of $\alpha$, which controls the relative variance
of the negative binomial degree distribution. The solid lines in figure \ref{mandT} are the theoretical results,
obtained from Eqs. (\ref{ja1}), (\ref{ja2}), and (\ref{hua1}), while the
symbols are numerical simulations of Eq. (\ref{jsa}) for large random graphs with $c=100$.
Due to the interplay between topological and thermal
fluctuations,  ${\rm Var}(m)$ is a non-monotonic function of $T$, with a maximum that shifts towards smaller temperatures for decreasing $\alpha$.
For $0 < \alpha \ll 1$, the degrees are very heterogeneous and a small amount of thermal noise leads to strong fluctuations of the local magnetizations.
Figure \ref{distrP} illustrates the typical shape of  $\mathcal{P}_{\rm h}(m)$ for a negative binomial degree distribution and different temperatures.
Similarly to the distribution of effective fields calculated in \cite{Metz2022}, the distribution $\mathcal{P}_{\rm h}(m)$
exhibits a power-law divergence at $m=0$ for $\alpha < 1$, which reflects the singular behavior of the rescaled degree distribution $\nu(g)$ at $g=0$.
The results in figures \ref{mandT} and  \ref{distrP} remain qualitatively the same for the algebraic distribution of the threshold noise.

In the case of the hyperbolic activation function $\mathcal{F}_{\rm h}(x)$, we
can expand Eqs. (\ref{ja1}) and (\ref{ja2}) in powers of $u$ for $0 < T_c -T \ll 1$ and show that
\begin{align}
  &m \simeq \pm   \sqrt{\frac{3 \langle G^2 \rangle}{ \langle G^4 \rangle  }}  \left( \frac{T_c - T}{T_c}  \right)^{\frac{1}{2}}, \label{jua1} \\
  & {\rm Var} (m)  \simeq \frac{3 \langle G^2 \rangle}{ \langle G^4 \rangle } \left(\langle G^2 \rangle - 1    \right) \left( \frac{ T_c - T  }{T_c} \right), \label{jua2}
\end{align}  
where $\langle G^n \rangle$ is the $n$-th moment of the distribution $\nu(g)$.
Consistently with previous works \cite{Doro2002,Leone2002,Dommers2016}, Eq. (\ref{jua1}) shows that $m$ exhibits the usual mean-field critical scaling when $\langle G^4 \rangle$ is finite.
The variance ${\rm Var} (m)$ vanishes linearly with $T_c - T$, analogously
to the variance of the replica-symmetric effective field distribution of fully-connected spin-glass models \cite{Sherrington1975}.

In the case of the algebraic activation function $\mathcal{F}_{\rm a}(x)$, an expansion in powers of $u$ contains
diverging coefficients, but we can still expand Eqs. (\ref{ja1}) and (\ref{ja2}) in powers of $u^{2 \kappa}$ and derive the
asymptotic behaviors
\begin{align}
  & m  \simeq \pm \left( \frac{2 \kappa \langle G^2 \rangle }{ \langle G^{2 \kappa+2} \rangle } \right)^{\frac{1}{2 \kappa}} 
  \left( \frac{T_c - T}{T_c}  \right)^{\frac{1}{2 \kappa}}, \label{tyes}  \\
  & {\rm Var} (m)  \simeq \left( \frac{2 \kappa \langle G^2  \rangle}{\langle G^{2 \kappa+2} \rangle }  \right)^{\frac{1}{\kappa}}
  \left(\langle G^2 \rangle - 1    \right)\left( \frac{ T_c - T }{T_c} \right)^{\frac{1}{\kappa}}.
  \label{tyes1}
 \end{align}
\begin{figure}[htp]
  \begin{center}
    \includegraphics[scale=0.58]{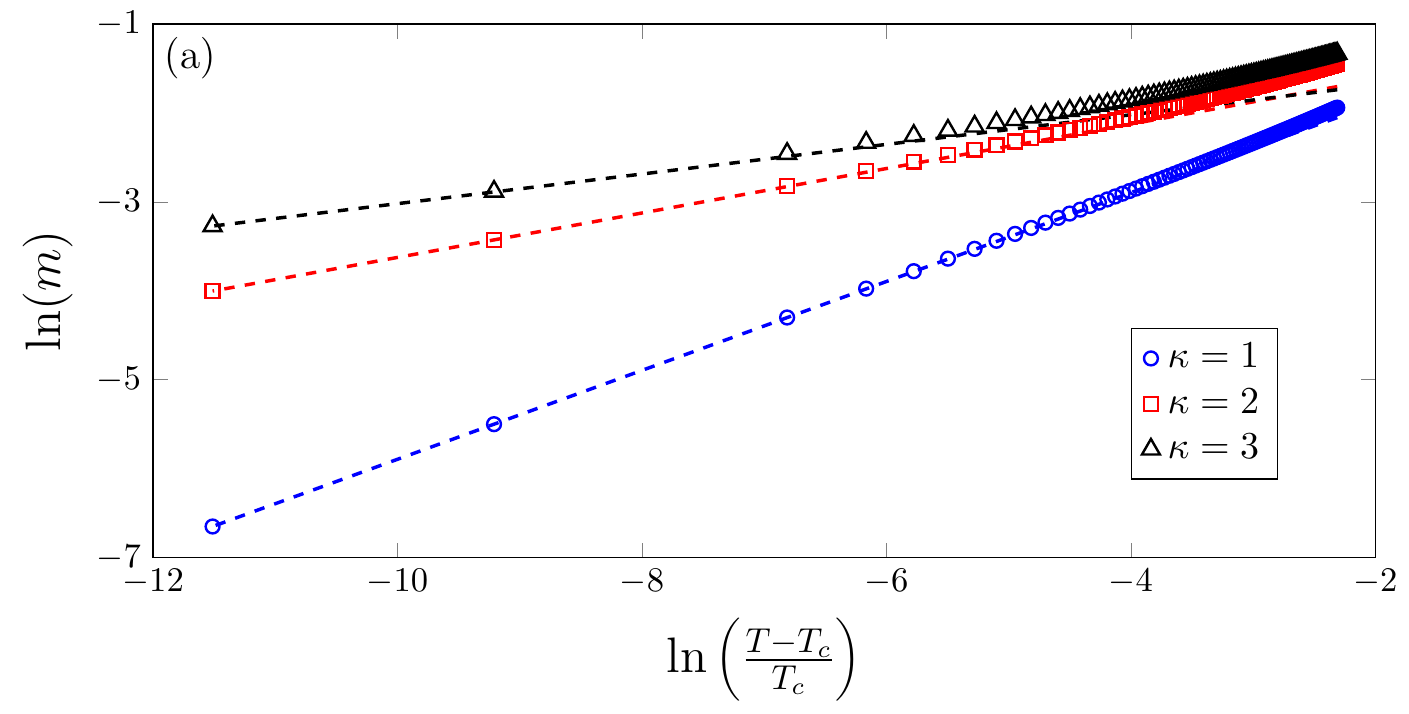} \\
    \hspace{-0.4cm}
     \includegraphics[scale=0.58]{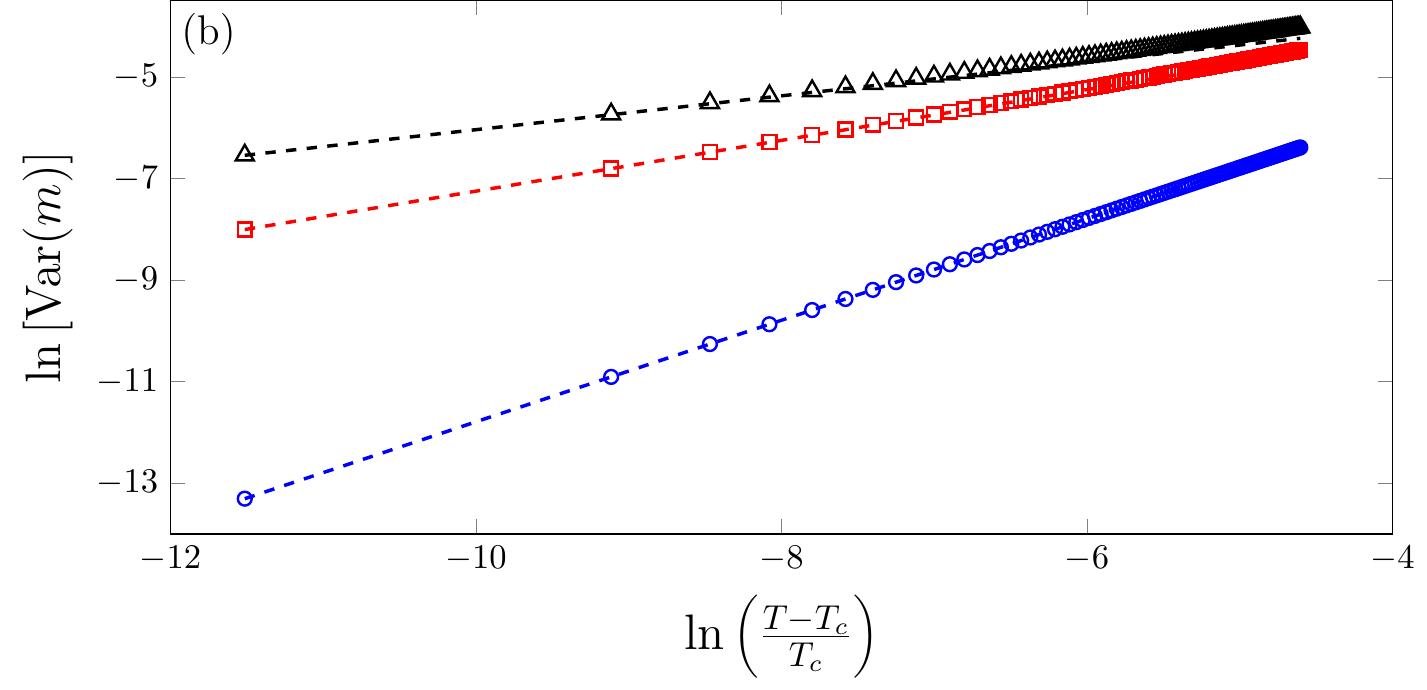}
     \caption{(a) The stationary magnetization $m$ and (b) the variance ${\rm Var}(m)$ of the distribution $\mathcal{P}_{\rm a}(m)$ of local magnetizations
       as a function of the reduced temperature $(T_c - T)/T_c$ (see Eq. (\ref{tutu})) for $\alpha=1$ and the algebraic
       distribution $\mu_{\rm a}$ of the threshold noise. The symbols are numerical results obtained from
       Eq. (\ref{hua2}), while the dashed lines are the analytic expressions of Eqs. (\ref{tyes}) and (\ref{tyes1}).
}
\label{idos}
\end{center}
\end{figure}
The above equations hold when $\langle G^{2 \kappa+2} \rangle$ is finite.
Remarkably, the critical exponents in Eqs. (\ref{tyes}) and (\ref{tyes1}) are determined by
the tails of the distribution $\mu_{\rm a}$ of the threshold noise. This is
a surprising finding for a mean-field model with long-ranged interactions between the spins.
Figure \ref{idos} compares Eqs. (\ref{tyes}) and (\ref{tyes1}) with numerical
solutions obtained from Eqs. (\ref{ja2}) and (\ref{hua2}) for different $\kappa$.
The agreement between the analytic results for the critical exponents and the numerical data is excellent.

%%%%%%%%%%%%%%%%%%%%%%%%%%%%%%%%%%%%%%%%%%%%%%%%%%%%%%%%%%%%%%%%%%%%%%%%%%%

\subsection{Dynamical behavior}

\begin{figure}[htp]
  \begin{center}
    \includegraphics[scale=0.6]{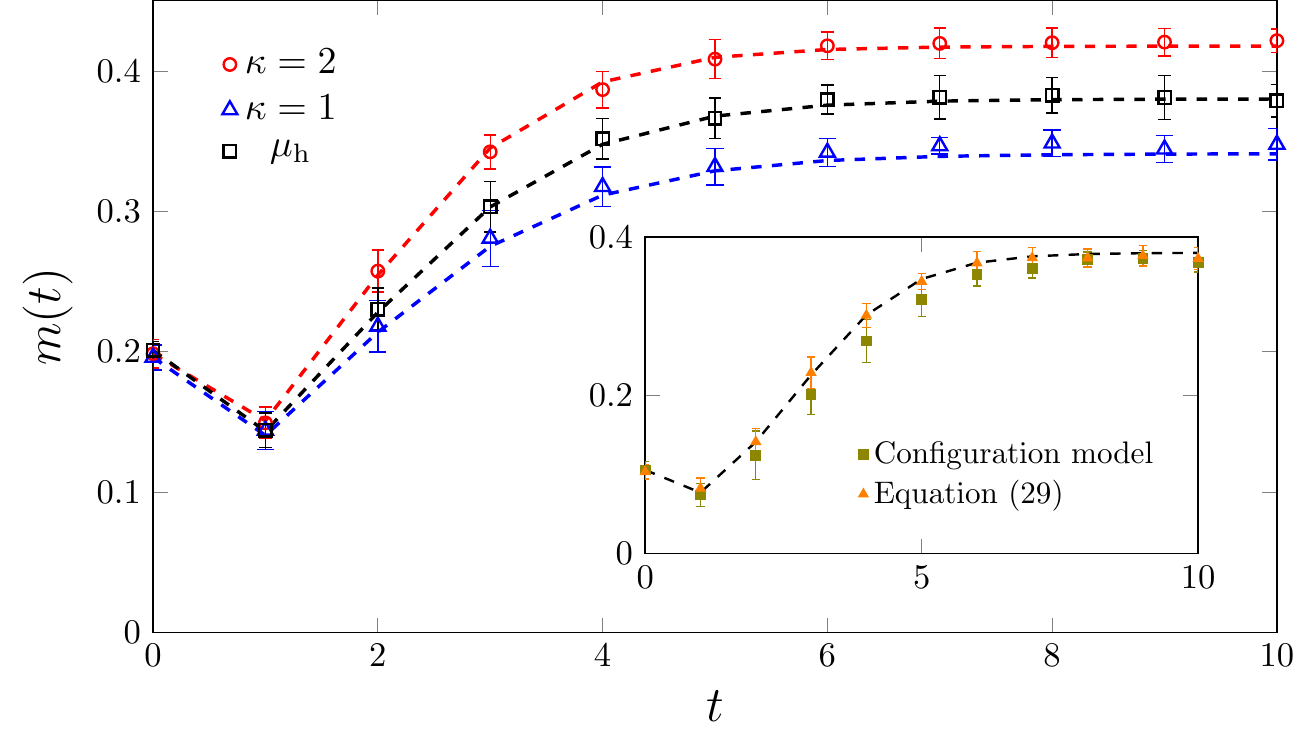}
    \caption{Dynamics of the average magnetization $m(t)$ inside the ferromagnetic phase of the Ising model on random graphs with
      a negative binomial degree distribution with $\alpha=0.625$.
      The dashed lines are derived from Eqs. (\ref{m1}) and (\ref{m2}), while
      the symbols are numerical simulations of Eq. (\ref{jsa}) for $N=10^4$ and mean degree $c = 100$.
      The main panel shows results for the hyperbolic tangent distribution and
      for the algebraic distribution (see Eq. (\ref{utq2})) of the threshold noise with temperature $T=T_{c}/2$. The random graphs in the numerical simulations
      of the main panel are sampled from Eq. (\ref{kda1}).
      The inset compares numerical simulation results obtained from the configuration model and from the effective matrix
      of Eq. (\ref{kda1}) for the distribution $\mu_{\rm h}$.
      The vertical bars are the standard deviations
      calculated from $10$ independent simulations.      
}
\label{compara}
\end{center}
\end{figure}

The nonequilibrium dynamics of the full distribution $\mathcal{P}(m,t)$ of local magnetizations is obtained
by iterating Eqs. (\ref{m2}) and (\ref{huhu1}) from an initial condition $u(0)$, which is related to the  local
magnetizations $m_1(0),\dots,m_N(0)$ by means of Eq. (\ref{bua1}).
Throughout this section we consider a homogeneous initial condition $m_{i}(0) = m(0)$ ($i=1,\dots,N$), which implies that $u(0)=m(0)$.
Figure \ref{compara} compares the iteration
of Eqs. (\ref{m1}) and (\ref{m2}) for the average magnetization $m(t)$ with numerical simulations of Eq. (\ref{jsa}) inside the ferromagnetic
phase, confirming the exactness of the theoretical recurrence equations for $c \rightarrow \infty$.
The finite-size simulation results of figure \ref{compara} converge for $t \gg 1$ to the fixed-point solutions obtained from Eqs. (\ref{ja1}) and (\ref{ja2}).
The inset of figure \ref{compara} shows numerical simulations on random graphs generated
through both the configuration model and the effective matrix of Eq. (\ref{kda1}) for the same system size, confirming
that the simulation results obtained from each method are approximately the same for large values of $N$ and $c$.

\begin{figure}[htp]
  \begin{center}
    \includegraphics[scale=0.6]{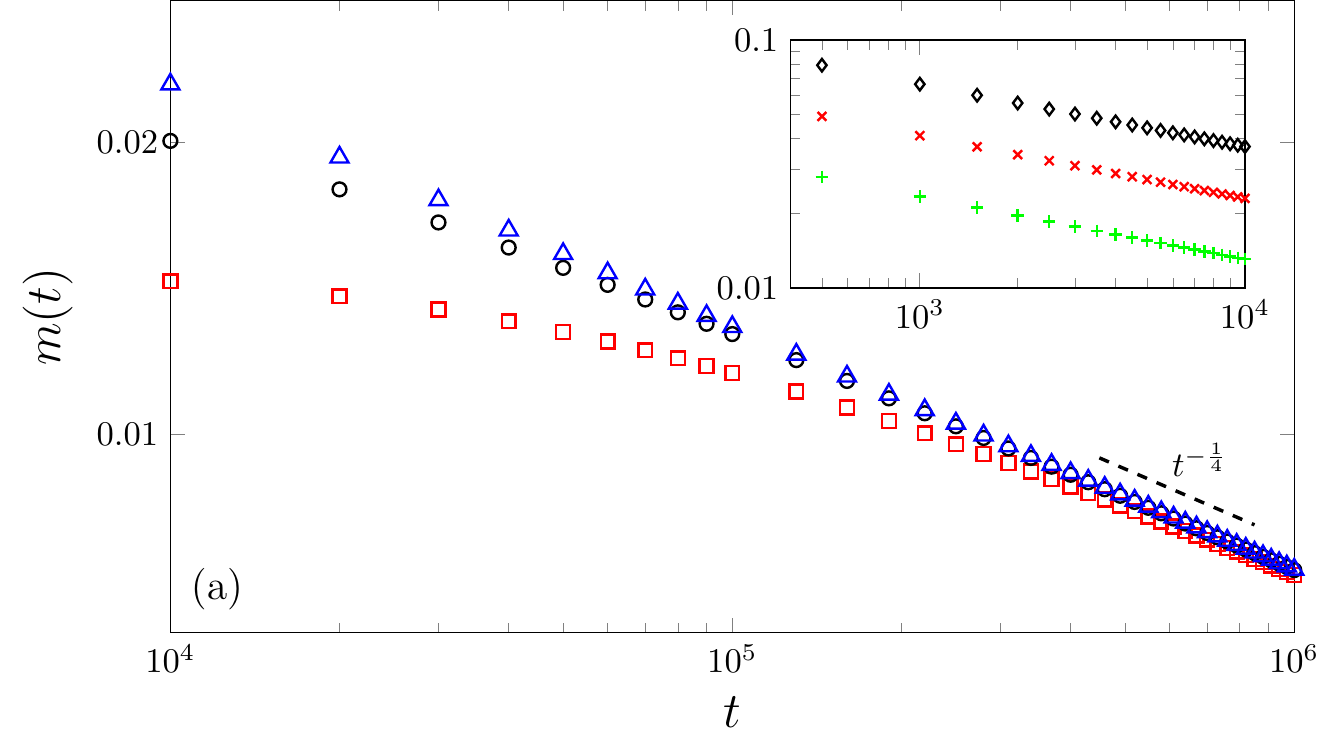} \\
    \hspace{-0.3cm}
    \includegraphics[scale=0.6]{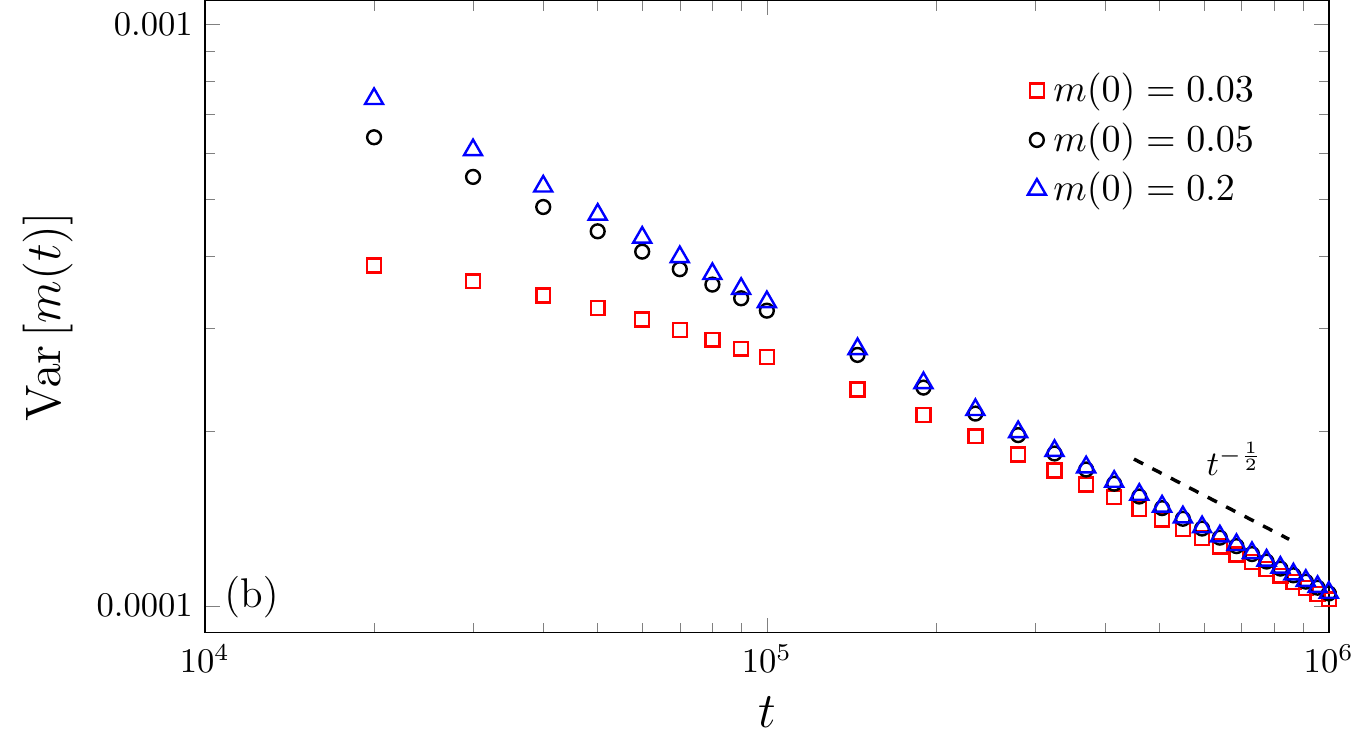}
    \caption{(a) Dynamics of the global magnetization $m(t)$ and of the (b) variance ${\rm Var} \left[ m(t) \right]$ of the distribution of local magnetizations
      at the critical temperature $T=T_c$. The degrees follow a negative binomial degree distribution with relative variance $1/\alpha$. The main panels show results for 
      $\alpha=1$, different initial conditions $m(0)$, and for the algebraic distribution $\mu_a(\zeta)$ of the threshold noise
      with $\kappa=2$. The inset in figure (a) illustrates the long-time power-law decay of $m(t)$ for $m(0)=0.4$ and different $\alpha$ (the other
      parameters are the same as in the main panels): $\alpha=2$ ($\diamond$), $\alpha=1$ ($\times$), and $\alpha=0.5$ ($+$). 
}
\label{dynTc}
\end{center}
\end{figure}

Now we discuss the nonequilibrium dynamics of $m(t)$ and ${\rm Var} \left[ m(t)  \right]$ at the critical temperature $T=T_c$.
After an initial transient that depends on $m(0)$, the average magnetization $m(t)$ and
the variance ${\rm Var} \left[ m(t)  \right]$ become independent of the initial conditions for large $t$ and they exhibit, respectively, the power-law decays
\begin{equation}
m(t) \propto \frac{1}{t^{z_{1}}} \quad \text{and} \quad {\rm Var} \left[ m(t)  \right] \propto \frac{1}{t^{z_{2}}},
\end{equation}  
with dynamical exponents $z_1$ and $z_2$ that only depend on the distribution of thresholds.
For the algebraic distribution $\mu_{\rm a}(\zeta)$ of the threshold noise, the dynamical exponents are given by $z_{1} = 1/2 \kappa$ and $z_{2} = 1/\kappa$, where
$\kappa$ controls the power-law tails of $\mu_{\rm a}(\zeta)$. Figure \ref{dynTc} illustrates the
critical dynamics of $m(t)$ and ${\rm Var} \left[ m(t)  \right]$ for $\kappa=2$.
For an hyperbolic tangent distribution $\mu_{\rm h}(\zeta)$, the exponents are given by $z_{1}=1/2$ and $z_{2}=1$. These are the standard mean-field values
for the critical dynamics of purely dissipative systems (models with nonconserved order-parameter) \cite{Janssen1989,Calabrese2005}.
Note that $z_1$ and $z_2$ have the same values as the critical indexes that govern the stationary critical behavior (see Eqs. (\ref{jua1}-\ref{tyes1})).
The inset of figure \ref{dynTc}-(a) clearly shows that $z_1$ and $z_2$  are independent of the variance $1/\alpha$ of the negative binomial degree distribution.

Lastly, we investigate how degree fluctuations and the distribution of thresholds
influence the dynamics of $m(t)$ inside each phase. 
For an arbitrary initial condition $0 < m(0) < 1$, the magnetization flows exponentially fast to its stationary state $m$, namely
\begin{equation}
  |m(t) - m| \propto e^{-t/\tau} \quad (t \gg 1).
  \label{judas}
\end{equation}
Close to a critical point, the relaxation time $\tau$ typically behaves as \cite{Odor2004}
\begin{equation}
\tau \propto \xi^Z,
\end{equation}  
where $\xi$ is the correlation length and $Z$ defines a dynamical exponent. In the homogeneous mean-field
Ising model, the correlation length and the relaxation time diverge, respectively, as $\xi \propto |T-T_c|^{-1/2}$ and
$\tau \propto |T-T_c|^{-1}$, which implies that $Z=2$ \cite{Odor2004}.
Below we examine
the critical scaling of $\tau$ in the present model.

\begin{figure}[htp]
  \begin{center}
    \includegraphics[scale=0.8]{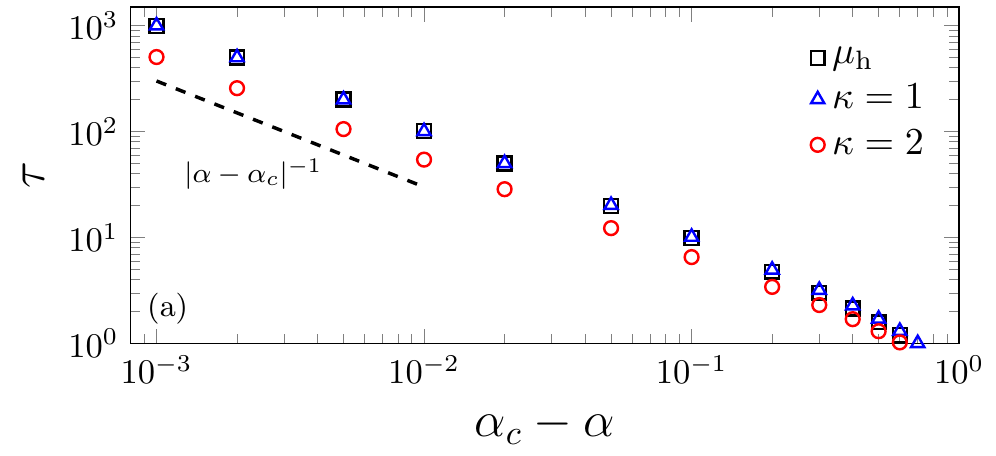} \\
    \includegraphics[scale=0.8]{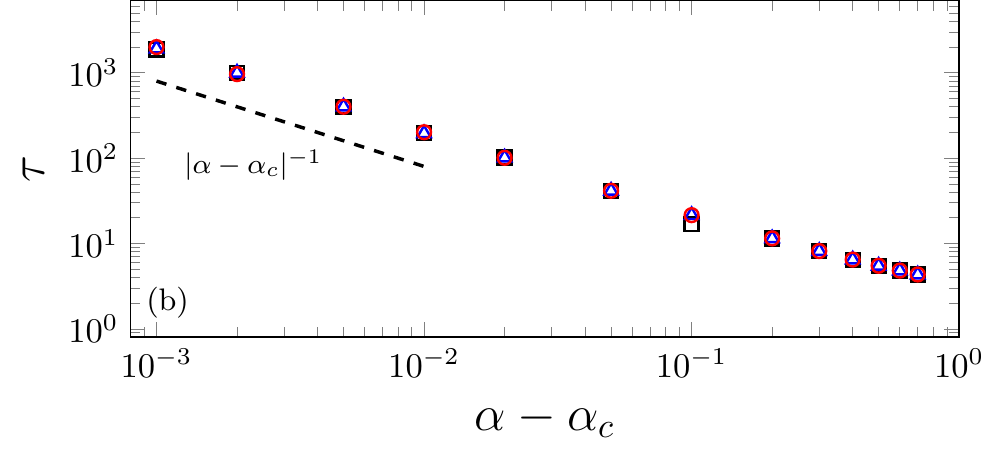}
    \caption{Relaxation time $\tau$ of the average magnetization as a function of $\alpha$ for fixed temperature $T= 2 J$, approaching the critical
      value $\alpha_c=1$ (see Eq. (\ref{tutu})) from the (a) ferromagnetic and from the (b) paramagnetic phase. The quantity $1/\alpha$
      is the relative variance of the negative binomial degree distribution. Each data point is obtained by fitting the long-time
      dynamics of $m(t)$, derived from the recurrence Eqs. (\ref{m1}) and (\ref{m2}), with the exponential function of Eq. (\ref{judas}). 
}
\label{TauFerro}
\end{center}
\end{figure}

Figure \ref{TauFerro} shows $\tau(\alpha)$ as a function of $\alpha$ for $T = 2 J$. As
the critical value $\alpha_c$ is approached from each side of the transition, the relaxation
time diverges as $\tau(\alpha) \propto |\alpha-\alpha_c|^{-1}$, independently of the distribution of thresholds.
For fixed $\alpha$, $\tau(T)$ also diverges as $\tau(T) \propto |T-T_c|^{-1}$, regardless the shape of  $\mu(\zeta)$.
Thus, it is reasonable to conclude that $Z=2$ in the present model, independently of the degree distribution
and of the distribution of thresholds.

%%%%%%%%%%%%%%%%%%%%%%%%%%%%%%%%%%%%%%%%%%%%%%%%%%%%%%%%%%%%%%
%%%%%%%%%%%%%%%%%%%%%%%%%%%%%%%%%%%%%%%%%%%%%%%%%%%%%%%%%%%%%%%%%

\section{Final remarks}
\label{sec:final}

We have presented an exact solution for the dynamics of the Ising model on highly-connected random graphs with an arbitrary degree distribution. The spins
are updated in parallel according to a stochastic dynamical rule which depends on a threshold noise that
emulates the contact of the system with a thermal bath.
For certain choices
of the distribution of thresholds, the microscopic stationary states of the dynamics do not follow the Boltzmann distribution, which rules out the
application of equilibrium statistical mechanics.

The solution of the model is given in terms of a general dynamical equation for the
distribution of local magnetizations, which
encapsulates all information about
the effect of both degree and 
threshold fluctuations in the behavior of the system.
The theoretical results for the stationary as well as for the nonequilibrium dynamics of the average magnetization have been validated by numerical simulations
of the microscopic dynamics. In addition, our numerical simulations have confirmed 
that the interaction matrix of the Ising model on graphs sampled from the configuration model
converges to the suitable matrix decomposition of Eq. (\ref{kda1}) in the high-connectivity limit. This equation enables to simulate the Ising model
on networks with an arbitrary degree distribution without resorting to more sophisticated algorithms to sample graph instances \cite{Fosdick_2018}.
We have shown that the model undergoes a continuous transition between a paramagnetic and a ferromagnetic phase.

We have presented results for random graphs with a negative binomial degree distribution, in which the high-connectivity limit
is solely parameterized by the variance of the rescaled degrees.
In particular, we have focused on the critical exponents that
characterize the stationary critical behavior and the long-time critical dynamics of the mean and the variance of the local magnetizations.
Our main result is to show that these critical exponents depend
on the distribution of the threshold noise.
If the distribution of thresholds is such that the model evolves to equilibrium states, then both exponents
assume their standard mean-field values \cite{Odor2004}. In contrast, if the distribution of thresholds is such that
the stationary states do not follow the Boltzmann distribution, then the aforementioned critical exponents may depend
on the fluctuations of the threshold noise. Remarkably, in the case of an algebraic threshold noise both
critical exponents are determined by the power-law tails of the distribution of thresholds.
In addition, we have shown that the dynamical exponent for the relaxation time of the average magnetization inside each phase
always assumes its standard mean-field value, regardless the distribution of the threshold noise.
Overall, our results show that the details of the microscopic dynamics and the absence of detailed balance are relevant factors in determining the
universality classes of spin models, in line with the critical properties of other nonequilibrium systems \cite{Odor2004}.
Still, the fact that the critical properties of the global magnetization do not belong to the mean-field
universality class is somewhat surprising, given that
random graphs can be  seen as the infinite dimensional, mean-field limit of finite-dimensional lattices.

Although the values of all critical exponents studied here are independent of the degree fluctuations, we point
out that all moments of the negative binomial degree distribution are finite. It is well established that, in the case of scale-free networks, the
equilibrium critical scaling of the magnetization depends on the power-law decay of the degree distribution \cite{Dommers2016} when its fourth
moment diverges. In this respect, it would be very interesting to consider the nonequilibrium dynamics of the Ising model
on scale-free networks and understand whether strong degree heterogeneities are able to modify the dynamical exponents of
spin models.

The probabilistic approach discussed here is general enough that it can be adapted to study the parallel as well as the sequential dynamics
of other agent-based models of interacting binary variables on undirected networks \cite{Castellano2009,Bouchaud2013,Torrisi2022,Hurry2022}. For sequential
dynamics, in which a single spin is updated at each time step, there is an extra critical index that describes 
the short-time, non-monotonic critical dynamics of spin models \cite{Janssen1989,Anteneodo2010}. Our work
paves the way to understand whether network heterogeneities change this critical exponent in mean-field models. We leave
this and other aforementioned problems as interesting perspectives of future works.

\begin{acknowledgments}

L. S. F. acknowledges a fellowship from CAPES/Brazil (finance code 001). F. L. M. thanks CNPq/Brazil for financial support.
  
\end{acknowledgments}

%%%%%%%%%%%%%%%%%%%%%%%%%%%%%%%%%%%%%%%%%%%%%%%%%%%%%%%%%%%%%%%%%%%%%%%%%%%%%%%%%%%%%
%%%%%%%%%%%%%%%%%%%%%%%%%%%%%%%%%%%%%%%%%%%%%%%%%%%%%%%%%%%%%%%%%%%%%%%%%%%%%%%%%%%%%%

\appendix

\section{Solution via the generating functional approach}
\label{app1}
In this appendix we present a more rigorous derivation of Eqs. (\ref{m1}) and (\ref{m2}) based on the generating
functional approach \cite{Martin1973,Coolen2001Dyn}. The time-dependent local field is defined by Eq. (\ref{loc}). We assume that
the entries $C_{ij}$ of the adjacency matrix $\boldsymbol{C}$ are generated according to
\begin{equation}
  C_{ij} = \frac{c}{N} G_i G_j (1-\delta_{ij}),
  \label{ghas}
\end{equation}
where $G_1,\dots,G_N$ are independent random variables drawn from the rescaled degree distribution $\nu(g)$. The above
effective matrix $\boldsymbol{C}$ has been put forward in section \ref{secder} based on the asymptotic form
of the local field for $c \rightarrow \infty$. Here we take Eq. (\ref{ghas}) as the definition of the adjacency
matrix elements and the starting point of our derivations, from which we will re-obtain the recurrence equation for the magnetization.
The calculations in this appendix further confirm that Eq. (\ref{ghas}) is the correct form of $\boldsymbol{C}$ for $c \rightarrow \infty$.

Our aim is to compute the disorder averaged generating functional
\begin{align}
  \mathcal{Z}[\boldsymbol{\psi}] =& \sum_{\boldsymbol{\sigma}(0),\dots,\boldsymbol{\sigma}(t)} \exp{\left[-i\sum_{s=0}^{t}  \sum_{j=1}^N \psi_j(s) \sigma_j(s)\right] } \nonumber  \\
  &\times p_{0}(\boldsymbol{\sigma}(0)) \left\langle  \prod_{s=0}^{t-1} W\left[\boldsymbol{\sigma}(s+1)|\boldsymbol{\sigma}(s)\right] \right\rangle_{\{ G_i \}},
  \label{popas}
\end{align}
where $p_{0}(\boldsymbol{\sigma}(0))$ is the probability distribution of the initial state and $\langle (\dots) \rangle_{\{ G_i \}}$ represents the average over the rescaled
degrees $G_1,\dots,G_N$.
The derivatives of $\mathcal{Z}[\boldsymbol{\psi}]$ with respect to the auxiliary fields $\{ \psi_i(t) \}$ yield all moments of the spin variables. For
instance, the magnetization follows from
\begin{equation}
  m(t) = \lim_{N \rightarrow \infty} \frac{i}{N} \sum_{i=1}^N \frac{\delta \mathcal{Z}}{\delta  \psi_i(t) } \Bigg{|}_{\psi=0} ,
  \label{gdaf}
\end{equation}  
where the shorthand notation $\psi=0$ means that $\psi_j(s)=0$ for any $j=1,\dots,N$ and $s=0,\dots,t$. 

By substituting Eq. (\ref{transitions}) in Eq. (\ref{popas}), we can rewrite $\mathcal{Z}[\boldsymbol{\psi}]$ as follows
\begin{align}
&\mathcal{Z}[\boldsymbol{\psi}] = \sum_{\boldsymbol{\sigma}(0),\dots,\boldsymbol{\sigma}(t)}  p_{0}(\boldsymbol{\sigma}(0)) \, e^{-i\sum\limits_{s=0}^{t}  \sum\limits_{j=1}^N \psi_j(s) \sigma_j(s) } \nonumber  \\
  &\times \int_{\mathbb{R}} \left( \prod_{j=1}^N \prod_{s=0}^{t-1} \frac{d h_j (s) d \hat{h}_j (s) }{4 \pi }
 \left[1 + \sigma_{j} (s+1) \mathcal{F} \left(\beta h_j (s)  \right)\right]  \right) \nonumber \\
 &\times e^{i \sum\limits_{j=1}^N \sum\limits_{s=0}^{t-1} h_j (s)  \hat{h}_j (s)   } \left\langle e^{- \frac{i J}{N} \sum\limits_{s=0}^{t-1} \sum\limits_{j k=1}^N   G_j G_k \hat{h}_j (s)   \sigma_k (s)    }
 \right\rangle_{\{ G_i \}},   \label{gugu1}
\end{align}  
in which the integration variables $\{ h_j (s) , \hat{h}_j (s) \}$ have been introduced through Dirac $\delta$-functions \cite{Coolen2001Dyn}.
In order to perform the average over $G_1,\dots,G_N$, we need to decouple sites in the exponent of the above equation, which
is achieved by inserting the macroscopic order-parameters
\begin{equation}
  u(s) = \frac{1}{N} \sum\limits_{i=1}^N G_i \sigma_i (s) \nonumber
\end{equation}
and
\begin{equation}
  v(s) = \frac{1}{N} \sum\limits_{i=1}^N G_i \hat{h}_i (s) \nonumber 
\end{equation}  
via Dirac $\delta$-functions that enforce the above definitions.
Moreover, by assuming that the initial states of the spins are independent, $p_{0}(\boldsymbol{\sigma}(0)) = \prod_{i=1}^N p_0(\sigma_i(0))$, the computation
of $\mathcal{Z}[\boldsymbol{\psi}]$ can be recast in terms of the calculation of an integral over the order-parameters
and its conjugate variables $\{ \hat{u}(s), \hat{v}(s) \}$, namely
\begin{equation}
  \mathcal{Z}[\boldsymbol{\psi}] = \int\limits_{\mathbb{R}} \left( \prod\limits_{s=0}^{t-1} \frac{N^2 d u(s) d v(s) d \hat{u}(s) d \hat{v}(s)  }{4 \pi^2}   \right) e^{N \Phi[u,v,\hat{u},\hat{v}]}.
  \label{iued}
\end{equation}  
The functional $\Phi[u,v,\hat{u},\hat{v}]$ is given by
\begin{align}
 \Phi[u,v,\hat{u},\hat{v}] & = i \sum\limits_{s=0}^{t-1} \left[u(s) \hat{u} (s) + v(s) \hat{v} (s) - J u(s) v(s)   \right] \nonumber \\
  &+ \frac{1}{N} \sum\limits_{j=1}^N \ln \Bigg{\{}  \sum\limits_{\vec{\sigma}}  \int_{\mathbb{R}} \frac{d \vec{h} d \vec{\hat{h}}} {(2 \pi)^t }
  e^{- i \sum\limits_{s=0}^t \psi_{j }(s) \sigma(s)}  \nonumber \\
  &\times \left\langle \mathcal{M}_G \left( \vec{h},\vec{\hat{h}},\vec{\sigma} \right) \right\rangle_G \Bigg{\}},
\end{align}  
where $d \vec{h} d \vec{\hat{h}} = \prod\limits_{s=0}^{t-1} d h (s) d \hat{h} (s)$ and
\begin{align}
  \mathcal{M}_{G} \left( \vec{h},\vec{\hat{h}},\vec{\sigma} \right) = p_0(\sigma(0))
  \prod_{s=0}^{t-1} \frac{1}{2} \left[1 + \sigma(s+1) \mathcal{F} \left( \beta h(s)  \right)   \right] \nonumber \\
  e^{i \sum\limits_{s=0}^{t-1} h(s) \hat{h}(s) - i G \sum\limits_{s=0}^{t-1}  \hat{u}(s) \sigma(s) - i G \sum\limits_{s=0}^{t-1}  \hat{v}(s) \hat{h}(s)    } .
\end{align}  
Note that we have defined the vectors $\vec{\sigma} = (\sigma(0),\dots,\sigma(t))^T$,  $\vec{h} = (h(0),\dots,h(t-1))^T$, and $\vec{\hat{h}} = (\hat{h}(0),\dots,\hat{h}(t-1))^T$, which
reflects the reduction of the computation of $\mathcal{Z}[\boldsymbol{\psi}]$ to an effective single-spin problem.

In the limit $N \rightarrow \infty$, the integral in Eq. (\ref{iued}) is solved by the saddle-point method and the generating functional
reads
\begin{equation}
  \mathcal{Z}[\boldsymbol{\psi}] \simeq \exp{\left(N \Phi_{*}[u,v,\hat{u},\hat{v}]   \right)},
  \label{hsgdf}
\end{equation}  
where $\Phi_{*}[u,v,\hat{u},\hat{v}]$ is the stationary value of $\Phi$. By deriving the functional $\Phi[u,v,\hat{u},\hat{v}]$
with respect to its arguments and then setting $\psi=0$, we obtain the saddle-point equations for the order-parameters and
their conjugate variables, 
\begin{align}
 & \hat{u}(l) = J v(l), \label{s1} \\
 & \hat{v}(l) = J u(l), \label{s2} \\
& u(l) = \frac{  \sum\limits_{\vec{\sigma}}  \int_{\mathbb{R}} d \vec{h} d \vec{\hat{h}}
 \, \sigma(l) \left\langle G \mathcal{M}_{G} \left( \vec{h},\vec{\hat{h}},\vec{\sigma} \right)  \right\rangle_G    }{  \sum\limits_{\vec{\sigma}}  \int_{\mathbb{R}} d \vec{h} d \vec{\hat{h}} 
   \left\langle \mathcal{M}_{G} \left( \vec{h},\vec{\hat{h}},\vec{\sigma} \right)  \right\rangle_G  } , \label{s3}  \\
  & v(l) = \frac{  \sum\limits_{\vec{\sigma}}  \int_{\mathbb{R}} d \vec{h} d \vec{\hat{h}} 
 \, \hat{h}(l) \left\langle G \mathcal{M}_{G} \left( \vec{h},\vec{\hat{h}},\vec{\sigma} \right)  \right\rangle_G     }{  \sum\limits_{\vec{\sigma}}  \int_{\mathbb{R}} d \vec{h} d \vec{\hat{h}}
 \left\langle \mathcal{M}_{G} \left( \vec{h},\vec{\hat{h}},\vec{\sigma} \right)  \right\rangle_G   }, \label{s4}
\end{align}  
which  give the arguments of $\Phi_{*}[u,v,\hat{u},\hat{v}]$. The magnetization 
\begin{equation}
m(t) = \frac{  \sum\limits_{\vec{\sigma}}  \int_{\mathbb{R}} d \vec{h} d \vec{\hat{h}}
  \, \sigma(t) \left\langle  \mathcal{M}_{G} \left( \vec{h},\vec{\hat{h}},\vec{\sigma} \right)  \right\rangle_G  }
{  \sum\limits_{\vec{\sigma}}  \int_{\mathbb{R}} d \vec{h} d \vec{\hat{h}}
  \left\langle \mathcal{M}_{G} \left( \vec{h},\vec{\hat{h}},\vec{\sigma} \right)  \right\rangle_G  }
\label{magg1}
\end{equation}
is obtained from Eqs. (\ref{gdaf}) and (\ref{hsgdf}).

The last step consists in simplifying the saddle-point equations. By adding a term of the form $G_j \theta_j(s)$ to the local field, Eq. (\ref{loc}), and then performing
the same calculation that led us to the above saddle-point integral, one finds that the single-site conjugate fields $\{ \hat{h}(s) \}$ couple 
to the external fields $\{ \theta_j(s) \}$ in such a way that the order-parameter $v(s)$ can be written in terms of the derivatives $\frac{\delta \mathcal{Z} }{\delta \theta_j(s)} \Big{|}_{\psi=0}$.
Combining this fact with the normalization property $\mathcal{Z}[0]=1$, one can show that $v(s)=0 \,\, \forall \, s$, which leads to the following
expression for the magnetization
\begin{align}
  m(t) &= \int_{\mathbb{R}} \frac{d \vec{h} d \vec{\hat{h}}} {(2 \pi)^t } \left \langle e^{i \sum\limits_{s=0}^{t-1} \hat{h} (s)   \left[  h(s) - G J u(s)  \right]  } \right \rangle_G  \nonumber \\
  & \times \sum_{\vec{\sigma}} p_0(\sigma(0)) \sigma(t)
  \prod_{s=0}^{t-1} \frac{1}{2} \left[1 + \sigma(s+1) \mathcal{F} \left( \beta h(s)  \right)   \right]
\end{align}  
and for the order-parameter $u(t)$
\begin{align}
  u(t) &= \int_{\mathbb{R}} \frac{d \vec{h} d \vec{\hat{h}}} {(2 \pi)^t } \left \langle G \, e^{i \sum\limits_{s=0}^{t-1} \hat{h} (s)   \left[  h(s) - G J u(s)  \right]  } \right \rangle_G  \nonumber \\
  & \times \sum_{\vec{\sigma}} p_0(\sigma(0)) \sigma(t)
  \prod_{s=0}^{t-1} \frac{1}{2} \left[1 + \sigma(s+1) \mathcal{F} \left( \beta h(s)  \right)   \right].
\end{align}  
Recalling that the random variable $G$ follows from the distribution $\nu(g)$, it is straightforward to recover Eqs. (\ref{m1}) and (\ref{m2}) by performing
the sum over $\vec{\sigma}$ and the integrals over
the fields in the above expressions.

\bibliography{biblio.bib}

\end{document}